# Наблюдение транзитного явления у звезды с экзопланетой WASP-2

*В.К. Игнатов[1], М.А. Горбачев[2,1], А.А. Шляпников[3]*

[1] ФГБУН «Крымская астрофизическая обсерватория РАН», Научный, Крым, Россия, 98409
*ivk@crao.ru*
[2] ФГАОУ ВО «Казанский (Приволжский) федеральный университет», Казань, Татарстан, Россия, 420000
*mark-gorbachev@rambler.ru*
[3] ФГБУН «Крымская астрофизическая обсерватория РАН», Научный, Крым, Россия, 98409
*aas@craocrimea.ru*

**Абстракт.** Представлены наблюдения транзита экзопланеты у звезды WASP-2, выполненные на телескопе МТМ-500 Крымской астрофизической обсерватории РАН. Рассмотрена краткая история открытия экзопланеты, её основные характеристики. Описаны наблюдения и процедура обработки. Выполнен анализ полученных результатов и проведено их сравнение с информацией из баз данных ETD и Архива экзопланет НАСА. В приложении к статье приведены данные фотометрии звёзды WASP-2 и звезды сравнения GSC 0052201406.

TRANSIT OBSERVATIONS OF THE EXOPLANET WASP-2B, *by V.K. Ignatov, M.A. Gorbachev and A.A. Shlyapnikov*. We present photometric transit observations of the exoplanet WASP-2b using the MTM-500 telescope of the Crimean Astrophysical Observatory of RAS. A brief history of the discovery of the exoplanet WASP-2b and its main characteristics are considered. Observations and the processing procedure are described. The analysis of the obtained results was carried out and compared with information from the ETD and NASA Exoplanet Archive databases. The photometry data of the star WASP-2 and the comparison star GSC 0052201406 are given in the supplementary table.

**Ключевые слова:** экзопланеты, фотометрия

## 1 Введение

После подписания в 2016 году нового договора о совместных исследованиях между Крымской астрофизической обсерваторией РАН и Институтом астрономии Национального университета Цинь Хуа (Тайвань), в КрАО проводятся систематические наблюдения звёзд с обнаруженными экзопланетами (Москвин и др., 2017). С целью поиска возможных пекулярных явлений во время транзита, в наблюдательную программу включены карлики нижней части главной последовательности, в том числе с активностью солнечного типа из каталога GTSh-10 (Гершберг и др., 2011).

В данной статье представлено наблюдение объекта WASP-2, которое было выполнено вне рамок договора с вышеуказанным Институтом астрономии. Полученные наблюдения, в том числе, дополняют базу данных фотометрических исследований звёзд в КрАО. Приведена информация об объекте, результаты выполненных наблюдений, проведен сравнительный анализ полученных данных с ранее опубликованными.

Наблюдение транзитного явления у звезды с экзопланетой WASP-2

## 2 WAS-2 b: обнаружение и основные параметры

WASP-2 двойная система, состоящая из карликовых компаньонов находящихся на расстоянии $0''.7$, спектральных классов K1.5 и M1-M4 (Бергфорс и др., 2013). Блеск системы в полосе $V$ составляет $11^m.98$, а показатель цвета $B - V$ равен $1^m.02$. Предположение о существовании экзопланеты в этой системе было сделано в рамках проекта SuperWASP (Стрит и др., 2003), запущенного в 2004 году. Подтверждением этой гипотезы стали дальнейшие наблюдения низкоамплитудных изменений лучевых скоростей, выполненных на спектрографе SOPHIE в Обсерватории Верхнего Прованса (Франция) в 2006 году. Обнаруженная в результате исследований экзопланета получила название WASP-2 b (Коллиер и др., 2007).

В таблице 1 представлены архивные данные об основных параметрах этой экзопланеты из работ (Торрес и др., 2008) и (Бономо и др., 2017).

Таблица 1

| Параметр | Единица измерения | Торрес и др., 2008 | Бономо и др., 2017 |
|---|---|---|---|
| Масса | $M_\oplus$ | $290.802^{+28.603}_{-29.557}$ | $294.0^{+25.7}_{-28.0}$ |
| Радиус | $R_\oplus$ | $12.005^{+0.897}_{-0.930}$ | $12.00^{+0.90}_{-0.93}$ |

## 3 Наблюдения WASP-2 в КрАО

В данной работе представлены наблюдения WASP-2, выполненные 2 июля 2018 г. на телескопе МТМ-500 (Менисковый телескоп Максутова, D/F = 500/6000 мм). В качестве детектора использовалась ПЗС-матрица Apogee U6 (2.4×2.4 см, 24 мкм/pix, 1024×1024 pix, $0''.72$/pix, FOV = $12'.4 \times 12'.4$). Наблюдения выполнялись в близкой к стандартной полосе $R_C$.

Результаты фотометрии были получены с помощью модернизированной программы потоковой обработки наблюдений (Москвин В.В., Шляпников А.А., 2017). В качестве звезды сравнения была выбрана GSC 0052201100. Результаты наблюдений приведены в фотометрическую систему каталога GAIA DR2 (GAIA Collaboration, 2018).

На рисунке 1 показаны кривые блеска звезды WASP-2, обозначенной как object, и GSC 0052201406, обозначенной как check star.

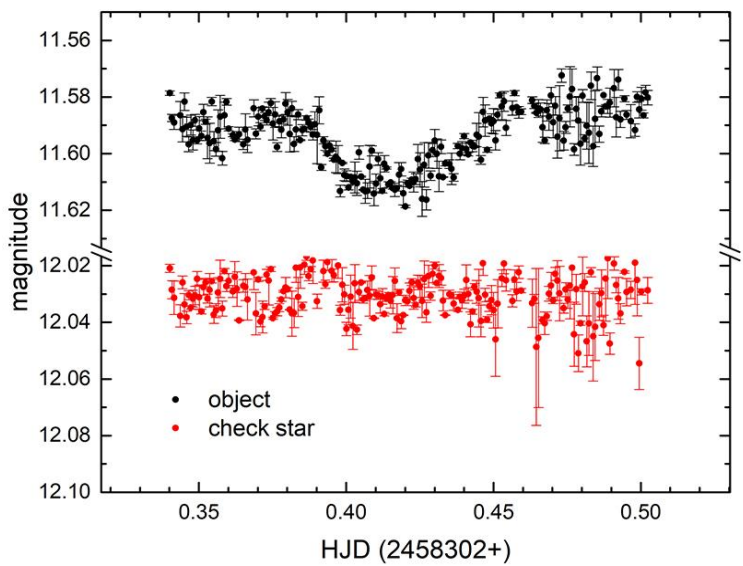

Рис. 1. Пояснения в тексте.

## 4 Анализ наблюдений в сравнении с информацией из сторонних баз данных

Для полученных результатов выполнен анализ с использованием таких баз данных как ETD (Брат и др., 2010) и Архив экзопланет НАСА[1]. База данных ETD предоставляет возможность аппроксимировать загружаемую кривую блеска транзитного явления, убирать, в случае

---

[1] Архив экзопланет НАСА - NASA Exoplanet Archive - https://exoplanetarchive.ipac.caltech.edu/index.html

В.К. Игнатов, М.А. Горбачев, А.А. Шляпников

необходимости, возникающий тренд (рис. 2) и представить результаты на диаграмме O – C (наблюдаемый момент времени середины транзита минус моделируемый). Это позволяет верифицировать некоторые исследуемые параметры экзопланетных систем.

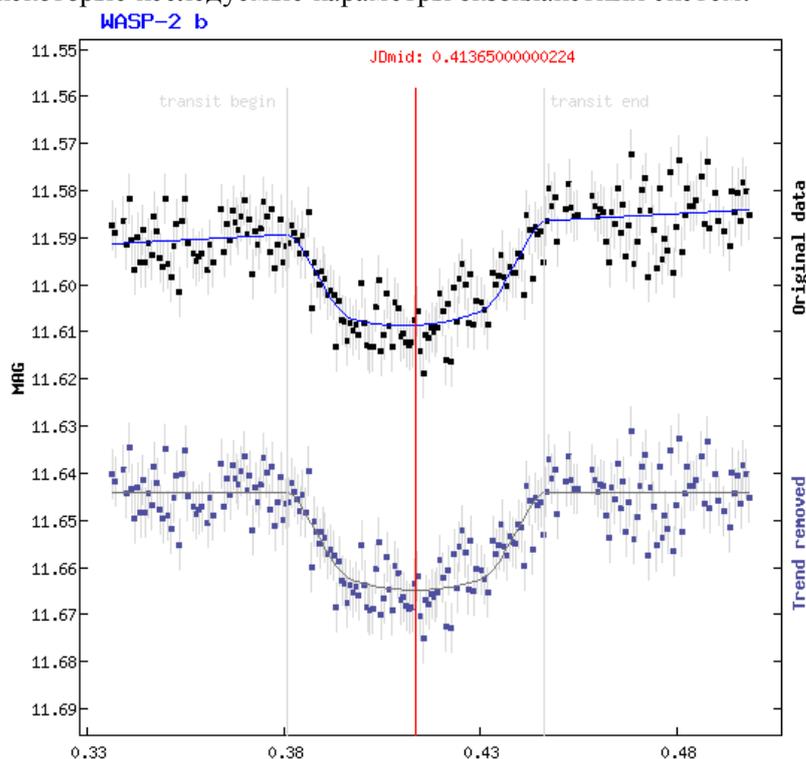

Рис. 2. Пояснения в тексте.

На рисунке 3 показана диаграмма O – C из ETD, где синим маркером обозначены представленные в данной работе наблюдения, а красными точками — наблюдения, размещённые в базе данных ранее.

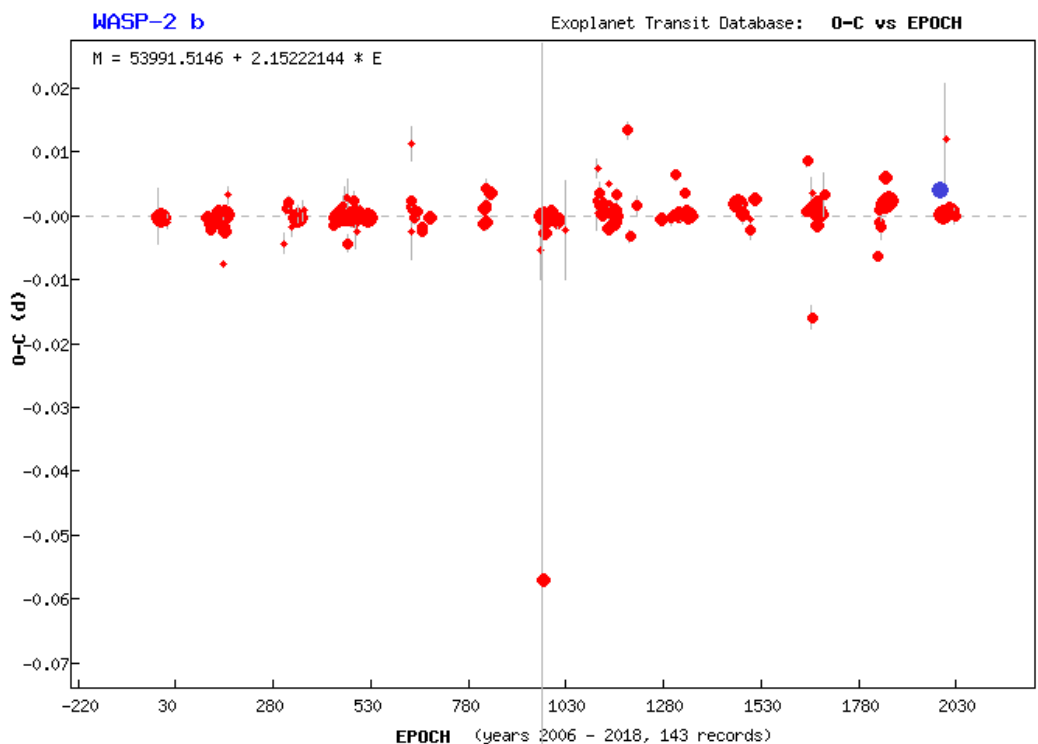

Рис. 3. Диаграмма O – C.

Аппроксимация кривой блеска транзитного явления в ETD позволяет определить такие параметры: момент времени середины транзита, глубину и продолжительность. Для анализа этих параметров были выбраны данные из Архива экзопланет НАСА.

Наблюдение транзитного явления у звезды с экзопланетой WASP-2

В таблице 2 показано сравнение параметров транзитного явления, определённых по представленным в данной статье наблюдениям и в опубликованных ранее работах (Триод и др., 2010 и Балуев и др., 2015).

Таблица 2

| Параметр | Единица измерения | Триод и др., 2010 | Балуев и др., 2015 | Данная статья |
|---|---|---|---|---|
| Момент времени середины транзита[*] | сутки | $2453991.51428^{+0.00020}_{-0.00021}$ | $2455894.07919 \pm 0.00015$ | $2458302.41824 \pm 0.00066$ |
| Глубина транзита | mag | $0.01802^{+0.00027}_{-0.00025}$ | — | $0.020728 \pm 0.000903$ |
| Продолжительность | минут | $106.14 \pm 0.96$ | $106.56 \pm 1.02$ | $94.1 \pm 2.7$ |

[*] гелиоцентрическая юлианская дата

## 5 Фотометрические данные

В приложении к статье представлены данные фотометрии звёзд WASP-2 и GSC 0052201406, которые продемонстрированы на рисунке 1. В первой колонке таблицы (*HJD*) указана гелиоцентрическая юлианская дата наблюдений в долях дня HJD = 2458302.0000000+. Во второй колонке (*R1*) приведены данные фотометрии WASP-2 и соответствующие ей ошибки в измерении звёздной величины (*errR1*). Четвёртая и пятая колонки содержат оценки блеска звёзд GSC 0052201406 (*R2*) и ошибки в его определении (*errR2*). Для большей компактности таблица размещена в статье в виде трех взаимодополняющих частей. Колонки 1 – 5 начало представления данных, 6 – 10 и 11 – 15, соответственно продолжение ряда фотометрических наблюдений.

## 6 Заключение

В данной работе представлено наблюдение транзитного явления у звезды с экзопланетой WASP-2, выполненное на телескопе МТМ-500 в КрАО. Показана кривая блеска транзитного явления, расположение полученных данных на диаграмме О – С, а также выполнен сравнительный анализ результатов наблюдений с данными из открытых источников.

В.К. Игнатов, М.А. Горбачев, А.А. Шляпников

*Приложение*

| HJD | R1 | errR1 | R2 | errR2 | HJD | R1 | errR1 | R2 | errR2 | HJD | R1 | errR1 | R2 | errR2 |
|---|---|---|---|---|---|---|---|---|---|---|---|---|---|---|
| 1 | 2 | 3 | 4 | 5 | 6 | 7 | 8 | 9 | 10 | 11 | 12 | 13 | 14 | 15 |
| .3362963 | 11.587 | 0.001 | 12.028 | 0.001 | .3805671 | 11.592 | 0.000 | 12.034 | 0.007 | .4204745 | 11.606 | 0.005 | 12.027 | 0.002 |
| .3370139 | 11.589 | 0.001 | 12.031 | 0.003 | .3812847 | 11.591 | 0.002 | 12.020 | 0.001 | .4211921 | 11.616 | 0.006 | 12.032 | 0.004 |
| .3391435 | 11.587 | 0.002 | 12.038 | 0.006 | .3819907 | 11.587 | 0.001 | 12.017 | 0.003 | .4218981 | 11.604 | 0.006 | 12.025 | 0.006 |
| .3398611 | 11.591 | 0.005 | 12.026 | 0.004 | .3827083 | 11.589 | 0.001 | 12.024 | 0.001 | .4226157 | 11.616 | 0.008 | 12.036 | 0.006 |
| .3405671 | 11.582 | 0.004 | 12.034 | 0.002 | .3834144 | 11.591 | 0.001 | 12.021 | 0.002 | .4233218 | 11.601 | 0.004 | 12.024 | 0.004 |
| .3412847 | 11.590 | 0.003 | 12.038 | 0.004 | .3841319 | 11.593 | 0.002 | 12.018 | 0.001 | .4240393 | 11.608 | 0.004 | 12.031 | 0.004 |
| .3419907 | 11.597 | 0.003 | 12.030 | 0.002 | .3848380 | 11.590 | 0.002 | 12.016 | 0.008 | .4247454 | 11.599 | 0.002 | 12.023 | 0.002 |
| .3427083 | 11.590 | 0.003 | 12.035 | 0.002 | .3855556 | 11.593 | 0.004 | 12.033 | 0.011 | .4254630 | 11.595 | 0.002 | 12.020 | 0.003 |
| .3434259 | 11.595 | 0.004 | 12.032 | 0.001 | .3862731 | 11.585 | 0.010 | 12.010 | 0.002 | .4261806 | 11.600 | 0.004 | 12.026 | 0.001 |
| .3441319 | 11.588 | 0.004 | 12.031 | 0.003 | .3869792 | 11.605 | 0.005 | 12.015 | 0.003 | .4268866 | 11.608 | 0.005 | 12.024 | 0.000 |
| .3448495 | 11.595 | 0.002 | 12.025 | 0.003 | .3876968 | 11.595 | 0.001 | 12.022 | 0.002 | .4276042 | 11.598 | 0.005 | 12.024 | 0.004 |
| .3455556 | 11.591 | 0.001 | 12.031 | 0.002 | .3884028 | 11.597 | 0.001 | 12.026 | 0.004 | .4283102 | 11.608 | 0.002 | 12.032 | 0.002 |
| .3462731 | 11.594 | 0.004 | 12.036 | 0.004 | .3891204 | 11.600 | 0.001 | 12.019 | 0.001 | .4290278 | 11.604 | 0.000 | 12.037 | 0.004 |
| .3469792 | 11.585 | 0.002 | 12.028 | 0.001 | .3898264 | 11.597 | 0.001 | 12.016 | 0.003 | .4304514 | 11.604 | 0.001 | 12.030 | 0.000 |
| .3476968 | 11.589 | 0.003 | 12.026 | 0.003 | .3905440 | 11.599 | 0.002 | 12.022 | 0.001 | .4311690 | 11.605 | 0.002 | 12.030 | 0.002 |
| .3484028 | 11.594 | 0.001 | 12.032 | 0.002 | .3912500 | 11.602 | 0.001 | 12.023 | 0.004 | .4318750 | 11.608 | 0.005 | 12.033 | 0.001 |
| .3491204 | 11.596 | 0.007 | 12.029 | 0.002 | .3919676 | 11.601 | 0.001 | 12.015 | 0.002 | .4332986 | 11.597 | 0.001 | 12.036 | 0.002 |
| .3498264 | 11.582 | 0.007 | 12.025 | 0.006 | .3926852 | 11.602 | 0.005 | 12.020 | 0.008 | .4340162 | 11.600 | 0.001 | 12.031 | 0.001 |
| .3505440 | 11.596 | 0.001 | 12.037 | 0.001 | .3933912 | 11.613 | 0.005 | 12.036 | 0.004 | .4347222 | 11.598 | 0.002 | 12.032 | 0.001 |
| .3512500 | 11.599 | 0.003 | 12.036 | 0.003 | .3941088 | 11.603 | 0.002 | 12.027 | 0.002 | .4354398 | 11.594 | 0.002 | 12.031 | 0.003 |
| .3519676 | 11.592 | 0.002 | 12.030 | 0.002 | .3948148 | 11.608 | 0.000 | 12.030 | 0.006 | .4361574 | 11.598 | 0.001 | 12.025 | 0.002 |
| .3526852 | 11.587 | 0.007 | 12.025 | 0.005 | .3955324 | 11.608 | 0.002 | 12.042 | 0.003 | .4368634 | 11.600 | 0.002 | 12.030 | 0.005 |
| .3533912 | 11.602 | 0.007 | 12.035 | 0.007 | .3962500 | 11.612 | 0.002 | 12.036 | 0.002 | .4375810 | 11.596 | 0.001 | 12.041 | 0.002 |
| .3541088 | 11.587 | 0.002 | 12.022 | 0.003 | .3969560 | 11.608 | 0.001 | 12.031 | 0.005 | .4382870 | 11.597 | 0.000 | 12.036 | 0.004 |
| .3548148 | 11.582 | 0.005 | 12.027 | 0.001 | .3976736 | 11.610 | 0.001 | 12.041 | 0.007 | .4390046 | 11.597 | 0.002 | 12.027 | 0.001 |
| .3555324 | 11.591 | 0.001 | 12.025 | 0.002 | .3983796 | 11.608 | 0.001 | 12.026 | 0.008 | .4397107 | 11.593 | 0.000 | 12.029 | 0.001 |
| .3569560 | 11.594 | 0.001 | 12.029 | 0.002 | .3990972 | 11.611 | 0.005 | 12.043 | 0.007 | .4404282 | 11.594 | 0.004 | 12.031 | 0.004 |
| .3576620 | 11.595 | 0.001 | 12.024 | 0.002 | .3998032 | 11.600 | 0.004 | 12.029 | 0.002 | .4411458 | 11.602 | 0.008 | 12.040 | 0.010 |
| .3583796 | 11.594 | 0.000 | 12.028 | 0.005 | .4005208 | 11.608 | 0.002 | 12.026 | 0.003 | .4418519 | 11.585 | 0.002 | 12.019 | 0.006 |
| .3590857 | 11.593 | 0.002 | 12.039 | 0.006 | .4012269 | 11.613 | 0.000 | 12.032 | 0.000 | .4425694 | 11.588 | 0.005 | 12.030 | 0.004 |
| .3605208 | 11.597 | 0.003 | 12.027 | 0.000 | .4019444 | 11.613 | 0.000 | 12.031 | 0.001 | .4432755 | 11.599 | 0.005 | 12.039 | 0.002 |
| .3612269 | 11.592 | 0.002 | 12.028 | 0.002 | .4026620 | 11.613 | 0.004 | 12.029 | 0.001 | .4439931 | 11.588 | 0.001 | 12.034 | 0.001 |
| .3619444 | 11.595 | 0.005 | 12.032 | 0.005 | .4033681 | 11.605 | 0.003 | 12.031 | 0.003 | .4446991 | 11.589 | 0.000 | 12.032 | 0.002 |
| .3640741 | 11.584 | 0.004 | 12.022 | 0.007 | .4040857 | 11.599 | 0.007 | 12.024 | 0.007 | .4454167 | 11.589 | 0.003 | 12.036 | 0.005 |
| .3647917 | 11.593 | 0.003 | 12.037 | 0.006 | .4047917 | 11.614 | 0.002 | 12.038 | 0.004 | .4461227 | 11.595 | 0.004 | 12.046 | 0.006 |
| .3656019 | 11.587 | 0.002 | 12.025 | 0.007 | .4055093 | 11.611 | 0.004 | 12.030 | 0.000 | .4468403 | 11.586 | 0.003 | 12.033 | 0.013 |
| .3663194 | 11.590 | 0.003 | 12.040 | 0.001 | .4062269 | 11.602 | 0.003 | 12.031 | 0.002 | .4475579 | 11.579 | 0.002 | 12.007 | 0.009 |
| .3670255 | 11.584 | 0.001 | 12.038 | 0.002 | .4069329 | 11.609 | 0.002 | 12.034 | 0.001 | .4482639 | 11.583 | 0.001 | 12.024 | 0.003 |
| .3677431 | 11.587 | 0.001 | 12.034 | 0.006 | .4076505 | 11.613 | 0.005 | 12.031 | 0.003 | .4489815 | 11.582 | 0.005 | 12.019 | 0.003 |
| .3684607 | 11.588 | 0.001 | 12.023 | 0.001 | .4083565 | 11.604 | 0.001 | 12.037 | 0.003 | .4496875 | 11.591 | 0.004 | 12.025 | 0.004 |
| .3691667 | 11.585 | 0.002 | 12.025 | 0.002 | .4090741 | 11.605 | 0.003 | 12.032 | 0.001 | .4518287 | 11.584 | 0.003 | 12.032 | 0.004 |
| .3698843 | 11.582 | 0.004 | 12.021 | 0.009 | .4097801 | 11.611 | 0.000 | 12.030 | 0.001 | .4525463 | 11.579 | 0.002 | 12.025 | 0.002 |
| .3705903 | 11.589 | 0.002 | 12.038 | 0.001 | .4104977 | 11.610 | 0.001 | 12.032 | 0.001 | .4532523 | 11.584 | 0.001 | 12.029 | 0.003 |
| .3713079 | 11.586 | 0.006 | 12.037 | 0.000 | .4112153 | 11.612 | 0.000 | 12.030 | 0.002 | .4539699 | 11.585 | 0.000 | 12.022 | 0.003 |
| .3720139 | 11.598 | 0.004 | 12.036 | 0.001 | .4119213 | 11.613 | 0.000 | 12.025 | 0.007 | .4547801 | 11.585 | 0.002 | 12.029 | 0.002 |
| .3727315 | 11.589 | 0.002 | 12.035 | 0.001 | .4126389 | 11.612 | 0.002 | 12.038 | 0.004 | .4585417 | 11.581 | 0.001 | 12.033 | 0.001 |
| .3734375 | 11.592 | 0.002 | 12.032 | 0.001 | .4133449 | 11.607 | 0.001 | 12.029 | 0.005 | .4592593 | 11.583 | 0.001 | 12.032 | 0.009 |
| .3741551 | 11.588 | 0.003 | 12.029 | 0.001 | .4140625 | 11.606 | 0.004 | 12.040 | 0.001 | .4599653 | 11.586 | 0.001 | 12.049 | 0.002 |
| .3748727 | 11.582 | 0.002 | 12.028 | 0.000 | .4147685 | 11.614 | 0.002 | 12.037 | 0.003 | .4606829 | 11.584 | 0.000 | 12.045 | 0.028 |
| .3755787 | 11.585 | 0.004 | 12.028 | 0.004 | .4154861 | 11.619 | 0.004 | 12.032 | 0.000 | .4613889 | 11.584 | 0.003 | 11.990 | 0.024 |
| .3762963 | 11.593 | 0.004 | 12.035 | 0.001 | .4162037 | 11.611 | 0.000 | 12.033 | 0.002 | .4621065 | 11.591 | 0.002 | 12.039 | 0.001 |
| .3770023 | 11.584 | 0.006 | 12.037 | 0.000 | .4169097 | 11.611 | 0.001 | 12.029 | 0.002 | .4628241 | 11.595 | 0.005 | 12.040 | 0.001 |
| .3777199 | 11.597 | 0.002 | 12.037 | 0.008 | .4176273 | 11.609 | 0.000 | 12.026 | 0.003 | .4635301 | 11.585 | 0.001 | 12.038 | 0.004 |
| .3784259 | 11.592 | 0.003 | 12.021 | 0.005 | .4183333 | 11.610 | 0.001 | 12.031 | 0.001 | .4642477 | 11.587 | 0.004 | 12.030 | 0.001 |
| .3791435 | 11.586 | 0.004 | 12.031 | 0.005 | .4190509 | 11.609 | 0.004 | 12.034 | 0.004 | .4649537 | 11.580 | 0.005 | 12.027 | 0.003 |
| .3798495 | 11.595 | 0.002 | 12.021 | 0.007 | .4197569 | 11.602 | 0.002 | 12.026 | 0.001 | .4656713 | 11.589 | 0.003 | 12.032 | 0.005 |

Наблюдение транзитного явления у звезды с экзопланетой WASP-2

*Приложение (продолжение)*

| *HJD* | *R1* | *errR1* | *R2* | *errR2* | *HJD* | *R1* | *errR1* | *R2* | *errR2* | *HJD* | *R1* | *errR1* | *R2* | *errR2* |
|---|---|---|---|---|---|---|---|---|---|---|---|---|---|---|
| *1* | *2* | *3* | *4* | *5* | *6* | *7* | *8* | *9* | *10* | *11* | *12* | *13* | *14* | *15* |
| .4663773 | 11.583 | 0.005 | 12.023 | 0.001 | .4763542 | 11.595 | 0.002 | 12.026 | 0.010 | .4870486 | 11.587 | 0.007 | 12.027 | 0.002 |
| .4670949 | 11.594 | 0.003 | 12.025 | 0.002 | .4770718 | 11.590 | 0.001 | 12.047 | 0.003 | .4877662 | 11.574 | 0.007 | 12.031 | 0.003 |
| .4678125 | 11.587 | 0.007 | 12.030 | 0.007 | .4777893 | 11.593 | 0.008 | 12.040 | 0.009 | .4884722 | 11.588 | 0.004 | 12.037 | 0.007 |
| .4685185 | 11.572 | 0.012 | 12.016 | 0.010 | .4784954 | 11.576 | 0.011 | 12.022 | 0.011 | .4898958 | 11.581 | 0.003 | 12.022 | 0.004 |
| .4692361 | 11.595 | 0.002 | 12.035 | 0.003 | .4792130 | 11.597 | 0.005 | 12.045 | 0.002 | .4906134 | 11.586 | 0.001 | 12.029 | 0.000 |
| .4699421 | 11.591 | 0.003 | 12.028 | 0.002 | .4799190 | 11.588 | 0.007 | 12.042 | 0.016 | .4920370 | 11.589 | 0.002 | 12.028 | 0.005 |
| .4706597 | 11.584 | 0.002 | 12.032 | 0.002 | .4806366 | 11.573 | 0.010 | 12.010 | 0.012 | .4934607 | 11.592 | 0.006 | 12.019 | 0.003 |
| .4713657 | 11.580 | 0.001 | 12.027 | 0.003 | .4813542 | 11.593 | 0.004 | 12.034 | 0.002 | .4941782 | 11.580 | 0.002 | 12.025 | 0.015 |
| .4720833 | 11.577 | 0.011 | 12.021 | 0.012 | .4820602 | 11.585 | 0.003 | 12.030 | 0.006 | .4948958 | 11.584 | 0.002 | 12.054 | 0.013 |
| .4727894 | 11.599 | 0.007 | 12.044 | 0.008 | .4827778 | 11.579 | 0.002 | 12.041 | 0.008 | .4956019 | 11.581 | 0.003 | 12.029 | 0.009 |
| .4735069 | 11.584 | 0.002 | 12.028 | 0.011 | .4834838 | 11.583 | 0.000 | 12.024 | 0.004 | .4963194 | 11.587 | 0.004 | 12.011 | 0.001 |
| .4742245 | 11.588 | 0.004 | 12.051 | 0.005 | .4842014 | 11.583 | 0.001 | 12.017 | 0.015 | .4970255 | 11.578 | 0.001 | 12.009 | 0.010 |
| .4749306 | 11.597 | 0.008 | 12.040 | 0.007 | .4849074 | 11.582 | 0.002 | 12.047 | 0.014 | .4977431 | 11.580 | 0.002 | 12.029 | 0.005 |
| .4756481 | 11.580 | 0.007 | 12.027 | 0.001 | .4863426 | 11.577 | 0.005 | 12.019 | 0.004 | .4984491 | 11.585 | 0.002 | 12.038 | 0.005 |